\documentstyle[aps,preprint,12pt,prl]{revtex}
\begin{document}
\draft
\title{2-Dimensional Electron Gas in a Linearly Varying Magnetic Field.
``Quantisation'' of the Electron and Current Density.}
\author{E. Hofstetter, J.M.C. Taylor and A. MacKinnon}
\address{Blackett Laboratory, Imperial College, London SW7 2BZ, UK}
\date{\today}
\maketitle
\begin{abstract}
We have developed new methods to calculate dispersion curves (analytically in
the simpler cases) from which we are able to derive the spatial distribution of
electron and current densities.  We investigate the case where the magnetic
field varies linearly with position and the results provide useful insights
into
the properties of this and other field distributions. We consider spin as well
as a confining electrostatic potential. We show that the electron and the
current density exhibit a very rich structure related to the quantisation of
the
energy. Moreover there is a direct contribution to the current density due to
the spin which could be of interest in relation to spin polarised current.
\end{abstract}
\vspace{1cm}
\section{Introduction}
A 2-dimensional electron gas (2DEG) in a magnetic field has proved to
be an extremely rich subject for theoretical and experimental
investigation \cite{Pran}. For example, considerable effort has been
devoted to the study of the Integral and Fractional Quantum Hall
Effects (I and FQHE), transport properties and edge states \cite{mst}.
Except in a few cases \cite{mul,chal}, however, the magnetic field
considered was homogeneous. In this article we address the problem of
a magnetic field varying linearly with position added to an underlying
homogeneous field. This is of relevance because in real systems: (a) a
constant magnetic field is not always attainable, and (b) an inhomogeneous
field may be desired. Another interesting point is that in the Composite
Fermion (CF) theory \cite{jai,chk1} which is used to describe the FQHE,
the electron-electron interaction, necessary for the appearance of the
FQHE, is incorporated into an effective magnetic field via a singular
gauge transformation. The result is a system of non-interacting
quasi-particles carrying a fictitious magnetic flux in an inhomogeneous
effective magnetic field. A better understanding of the properties of a
simple non-interacting electron gas in an inhomogeneous field might
therefore bring useful insights into CF theory.
\section{Model}
To investigate the electronic properties of a non-interacting electron
gas in a linearly varying magnetic field we consider the following Hamiltonian
\begin{equation}
H=\frac{1}{2m^{\ast}}\left({\bf p}-
e{\bf A}\right)^{2}-\frac{{\rm g} e}{2 m_{e}}{\bf SB} +V_{c},
\end{equation}
where ${\bf A}=(\frac{1}{2}B_{1}y^{2}+B_{0}y,0,0)$,
${\bf B}=-(0,0,B_{1}y+B_{0})$, the confining potential due to the walls
$V_{c}(y)=\beta \left[\exp\left(\frac{\alpha}{y_{e}}(y-y_{e})\right)+
\exp\left(-\frac{\alpha}{y_{e}}(y+y_{e})\right) \right]$, $y_{e}$ is
the position of the edge of the system, $\bf S$ is the spin operator and
${\rm g}=2$ for the g-factor. The parameters $\alpha$ and
$\beta$ allow the shape of the potential to change continuously from very
sharp to very smooth, which can modify the properties of the system
\cite{chk1,chk2}. By solving
\begin{equation}
\left(\frac{1}{2m^{\ast}}\left({\bf p}-
e{\bf A}\right)^{2}-\frac{e}{m_{e}}{\bf SB}+V_{c}\right)
\chi(x,y)= E \chi(x,y)
\end{equation}
we can obtain the electron density
\begin{equation}
\rho(x,y)=\sum_{{\rm states}}\ \chi^{*}(x,y)\chi(x,y),
\end{equation}
where we sum over all states with energy $E\leq E_{F}$, and the current
density for a state $n$ \cite{lan}
\begin{equation}
{\bf j}^{(n)}(x,y)=\frac{e}{m^{\ast}}{\rm Re}\left[ \chi^{\ast}_{n}(x,y)
\left({\bf p}- e{\bf A}\right)
\chi_{n}(x,y)\right]+
\frac{e}{m_{e}}\nabla\times\left[
\chi^{\ast}_{n}(x,y){\bf S}\chi_{n}(x,y)\right].
\end{equation}
As a result of our choice of ${\bf A}$ and ${\bf B}$, the
symmetry of the Hamiltonian allows us to write the
wave function as $\chi(x,y)={\rm e}^{ik_{x}x}\psi(y)$, where $k_{x}$ is
a good quantum number, and we then obtain the 1-dimensional Schr\"{o}dinger
equation
\begin{equation}
\left[\frac{p_{y}^{2}}{2m^{\ast}}+\frac{1}{2m^{\ast}}\left(\hbar k_{x}-
eA_{x}\right)^{2}-\frac{e \hbar s}{m_{e}}B_{z}+V_{c}\right] \psi(y) =
E \psi(y),
\end{equation}
with $s=\pm 0.5$.
Eq.(5) then enables us to derive the dispersion curves $E_{n}(k_{x})$,
and to rewrite the electron density as
\begin{equation}
\rho(y)=\sum_{n} \sum_{k_{x}} \psi^{*}_{n}(y)\psi_{n}(y)
\end{equation}
and the current density for one state $n$ and a fixed $k_{x}$ as
\begin{equation}
j^{(n)}_{x}(y)=\frac{e}{m^{\ast}}\left[ \psi^{\ast}_{n}(y)
\left(\hbar k_{x}-e A_{x}\right)
\psi_{n}(y)\right]+
\frac{e \hbar s}{m_{e}}
\frac{\partial\left[\psi^{\ast}_{n}(y)\psi_{n}(y)\right]}{\partial y},
\end{equation}
both now solely functions of $y$. Due to the symmetry of the system there is a
current density only along the $x$ axis. Integrating over
$y$ will give the total current $I_{x}^{n}$ carried, for a fixed $k_{x}$ by
the state $n$ and summing over $n$ and $k_{x}$ the total current $I_{x}$.
\section{Method}
Starting from Eq.(5) the Hamiltonian can be written as
\vspace{0.1cm}
\begin{eqnarray}
\left[\frac{p_{y}^{2}}{2m^{\ast}}+\frac{m^{\ast}\omega_{1}^{2}}{8}y^{4}+
\frac{m^{\ast}\omega_{1}\omega_{0}}{2}y^3+\left(\frac{m^{\ast}\omega_{0}^{2}}{2}
-\frac{\hbar k_{x}\omega_{1}}{2}\right)y^2+ \right.
\nonumber
\\
\left.\left(\frac{e \hbar s B_{1}}{m_{e}}-
\hbar k_{x}\omega_{0}\right)y+\left(\frac{e \hbar s B_{0}}{m_{e}}
+\frac{\hbar^{2}k_{x}^{2}}{2m^{\ast}}\right)+V_{c}(y)\right]
\psi(y)&=&E\psi(y),
\end{eqnarray}
\vspace{-0.4cm}
\\
with $\omega_{0}=\frac{eB_{0}}{m^{\ast}}$ and
$\omega_{1}=\frac{eB_{1}}{m^{\ast}}$. For what follows it is useful to
introduce the dimensionless variable
$\hat{y}=\left(\hbar^{-1}m^{\ast}\omega_{1}\right)^{\frac{1}{3}}y$
and $\hat{p}=\left(m^{\ast}\omega_{1}\hbar^{2}\right)^{-\frac{1}{3}}p$ which
yields for (8)
\begin{equation}
\left(\frac{p_{\hat{y}}^{2}}{2m'}+a\hat{y}^{4}+
b\hat{y}^{3}+c\hat{y}^{2}+d\hat{y}+e+
V_{c}(\hat{y})\right)\psi(\hat{y}) = E \psi (\hat{y}),
\end{equation}
with $m'=\frac{m^{\ast}}{(m^{\ast}\omega_{1}\hbar)^{2/3}}$ and $a,b,c,d,e$
now given in unit of energy.
Although in some simplified cases it is
possible to obtain analytical results, as we will see below, there is in
general no way to find the analytical solution of Eq. (9), and therefore we
have to resort to numerical calculations. Eq.(9) can be solved by expanding
$\psi(y)$ in
terms of oscillator functions, $\phi_{n}(\hat{y})={\rm H}_{n}(\hat{y})
{\rm e}^{-\frac{\hat{y}^{2}}{2}}$ where ${\rm H}_{n}$ is a Hermite polynomial,
and then
by numerically diagonalising the corresponding secular equation
\begin{equation}
{\rm Det}|H_{kn}-E\delta_{kn}|=0.
\end{equation}
Using the properties of the Hermite polynomials all the matrix elements
$\langle\phi_{k}|H|\phi_{n}\rangle$ can be calculated analytically (Appendix)
which greatly improves the diagonalisation method. However before starting with
the numerical calculations we can try an analytical approach to Eq.(8) in the
simplified case where $B_{0}=0$ and $V_{c}=0$. We then have
\begin{equation}
\left[\frac{p_y^2}{2m^*}+\frac{\hbar^2}{2m^*}\left(\frac{e^2B_1^2}
{4\;\hbar^2}y^4-\frac{eB_1}{\hbar}k_xy^2+\frac{2eB_1m^*s}{\hbar
m_e}y+k_x^2\right)\right]\psi(y)=E \psi(y).
\end{equation}
We choose two regimes for which $B_1\neq 0$: (a)
$k_x<0$, single well potential (SWP), near $B=0$, and (b) $k_x>0$, double
well potential (DWP), near $B=\pm B_1\left|\sqrt{2\hbar k_x/eB_1}\right|$. We
expand parabolically around the minima of the effective potential and obtain
harmonic oscillator equations which we solve analytically. The expressions
obtained for the energy are
\begin{equation}
{\rm SWP}:\hspace*{0mm}E_{n}=\frac{\hbar^2}{2m^*}
\left[k_x^2+(2n+1)\sqrt{\frac{k_x eB_1}{\hbar}}-
\frac{ eB_1 \eta^{2}}{\hbar k_x}\right]
\end{equation}
\begin{equation}
{\rm DWP}:\hspace*{0mm}E_{n}=\frac{\hbar^2}{2m^*}\left[(2n+1)
\sqrt{\frac{2k_x eB_1}{\hbar}}-\frac{eB_1\eta^2}
{2\hbar k_x}\pm 2 \eta \sqrt{\frac{2k_x eB_1}{\hbar}}\right],
\end{equation}
where $\eta=\frac{m^*s}{m_e}$. From here it is straightforward to derive the
group velocity ($\frac{1}{\hbar}\frac{dE_{n}}{dk_{x}}$) for the state $n$ as
\begin{equation}
{\rm SWP}:\hspace*{0mm}v_x=\frac{\hbar}{m^*}\left[k_x+\frac{eB_1\eta^2}
{2\hbar k_x^2}+(2n+1)
\sqrt{\frac{eB_1}{16\hbar k_x}}\right]
\end{equation}
\begin{equation}
{\rm DWP}:\hspace*{0mm}v_x=\frac{\hbar}{m^*}\left[\frac{eB_1\eta^2}
{4\hbar k_x^2}+(2n+1)\sqrt{\frac{eB_1}{8\hbar k_x}}\pm\eta\sqrt{\frac{eB_1}
{2\hbar k_x}}\right]
\end{equation}
\section{Results}
In the following calculations the system we consider corresponds to an ideal
slab of $\rm{GaAs/Al_xGa_{1-x}As}$ heterostructure filled with an ideal 2DEG.
The effective electron mass is $m^*=0.067\ m_e$, the electron density
$4\times 10^{-5} {\it \AA}^{-2}$, and  the sample has width, when $V_{c}\neq
0$,
$2y_e=2\times 10^4{\it \AA}$ and length $L\gg 1$ with periodic boundary
conditions along $x$.
\par
In Fig.\ref{1} are reported the numerical and the analytical results
(Eqs.(12),(13)) for the case when $V_{c}=0$, $B_{1}=1 G/{\it \AA}$, and
$B_{0}=0$. We can see that the agreement is very good except around $k_{x}=0$
where the method breaks down. Although it is possible to obtain
useful information from analytical calculations they do not allow us to derive
complete dispersion curves and hence the electron or current densities.
Moreover we are interested in taking into account the effects of a confining
potential $V_{c}$ but this cannot be included in our analytical approach. We
have then to use a numerical approach.
\par
So, starting from the same model system configuration as above but this time
with $V_{c}\neq 0$ and, $B_{0}=0.5*10^{4} \ G$.
We chose $\alpha=100$ and $\beta=50$ which correspond to quite a sharp
confining potential. The dispersion curves are plotted in Fig. \ref{2}.
The degeneracy of the energy levels is completely removed and the new structure
appearing in the dispersion curves is due to the breaking of the $y$ symmetry
in
the Hamiltonian. It is interesting to note that some similar features were
observed in the case of a curved 2DEG in a constant magnetic field \cite{fod}.
Having $E_{n}$ and $\psi_{n}$ we can now calculate $j_{x}^{(n)}(y)$ and
$\rho(y)$,
but to do this we need the Fermi energy $E_{F}$. This can be obtained by
minimising the total energy with the constraint that the number of electrons
$N$
is constant with $N$ given by
\begin{equation}
N=\frac{1}{2 \pi}\sum_{n}\int_{k_{x1}^{(n)}}^{k_{x2}^{(n)}} dk^{(n)}_{x},
\end{equation}
where $\{k_{xi}^{(n)}\}$ are the parameters we vary to minimize the energy.
As we might expect, $E_{F}$ does not depend on $k_{x}$ but, in contrast to the
assumption in \cite{mul}, it is not independent of the magnetic field. This
is shown in Fig. \ref{3} where $E_{F}$ has been calculated for various values
of $B_{1}$ and $B_{0}$. This dependence is due, in the absence of external
leads, to the walls, which can be seen by the fact that when $B_{1}$, which in
contrast to $B_{0}$ removes the degeneracy of the states and gives rise to an
effective confining potential, increases, $E_{F}$ becomes independent of the
magnetic field.
\par
Using $E_{F}$ we can now calculate the electron density $\rho(y)$. In
Fig. \ref{4} $\rho(y)$ is plotted for different values of $B_{1}$ and $B_{0}$.
We see that with an external confining potential ($V_{c}\neq 0$) $\rho(y)$
is not constant; it can have quite a rich structure with local charging
effects.
Moreover, when $B_{0}\neq 0$, $\rho(y)$ becomes asymmetric in $y$. It has to be
noted that in the case $V_{c}=0$ $\rho(y)$ is constant the oscillations on left
and right hand side of the graphic are only a numerical effect due to the fact
that in this case one should use a larger set of $\phi_{n}$ for
the expansion of $\psi$. But for $V_{c}\neq 0$ the structure of $\rho(y)$ can
be explained in the following way. When the number of states as a function of
$k_{x}$ is plotted a discontinuous curve due to the quantisation of the energy
is obtained. Although there is no simple relation between the $k_{x}$ space
and the real space, as for a homogeneous magnetic field, the number of
oscillations of the electron density is the same as the number of steps in the
discontinuous curve and then is indeed a consequence of the energy
quantisation.
\par
\begin{figure}
\caption{Same as Fig. 11 but with $B_{1}= 3 \  G / {\it\AA}$, $B_{0}=2 \times
10^{4} \ G$.}
\label{12}
\end{figure}
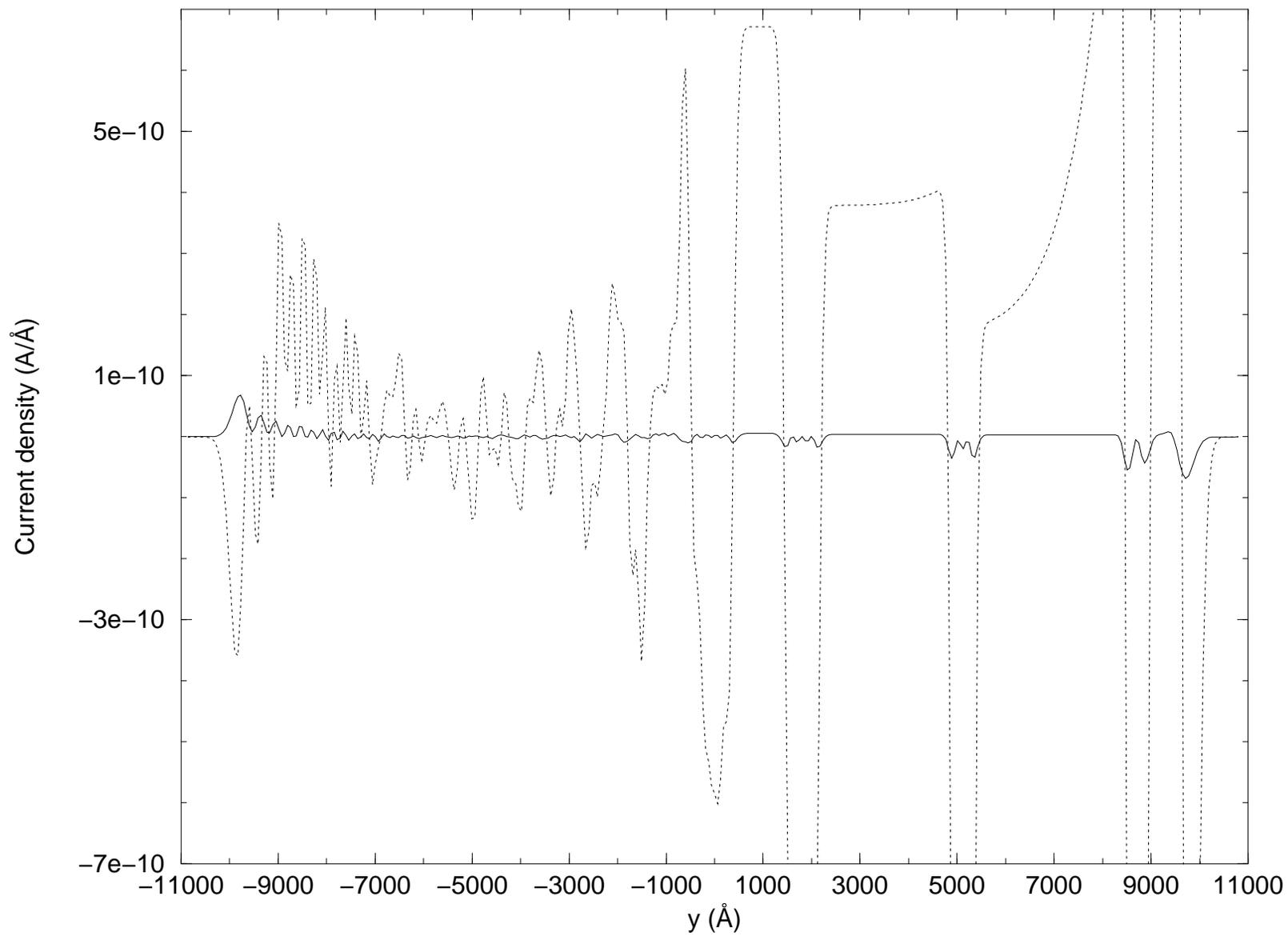

This can be seen in Figs. \ref{5} and \ref{6}. The next step now is, by means
of Eq.(7), to calculate the total current density given by
\begin{equation}
j_{x}(y)=\sum_{{\rm states}} j_{x}^{(n)}(y),
\end{equation}
where the sum runs over all the states with $E\leq E_{F}$. In Fig. \ref{7} are
reported the current densities for different magnetic fields
disregarding first the part due to the spin. One sees that the
shape of the current density is much more subtle than might be expected
from semi-classical approximations. Actually considering $j_{x}^{(n)}(y)$ for
$|k_{x}|\gg 0$ or, in other words for the largest $k_{x}$ close to the Fermi
energy,
the movement of the electron can be described by its classical
orbits (drifting orbits along the edge and snake orbits in the opposite
direction along the line where $B\simeq 0$). For smaller $k_{x}$, however, the
states tunnelling between the two wells but mainly the ones of energy above the
central maximum of the double well are very important and their contribution
cannot be overlooked. In fact $j_{x}^{(n)}(y)$, after summing over all $n$ and
$k_{x}$, turns out to be very small for the case where $B_{0}=0\ {\rm for\
all}\
y$.
In Fig. \ref{8} are reported different $j_{x}^{(n)}(y)$. For
$k_{x}= -0.018 {\it \AA}^{-1}$, $k_{x}=0.08 {\it \AA}^{-1}$ and $n=0$ the
movement of the
electron is well defined by its classical orbits (edge and snake orbits). But
around $k_{x}=0$ and for exemple here $n=31$ the situation is more complicated.
An
interesting point is that now the current density for one state can be positive
{\it and} negative as a function of $y$. Moreover we see that the positive part
is located in $y$ where the density of current flows in the other direction due
to the presence of the snake orbits. The same kind of phenomena appears with
edge orbits. This can be understand by considering the first term
$\hbar k_{x}- \frac{1}{2} e B_{1} y^{2}- e B_{0} y$
in Eq.(7). It is easy to imagine that when summing the current density over all
the different states the result is rather different from what we might expect
from the consideration of the simple classical picture of the orbits. It has
to be noted that although the current density can be positive and negative as a
function of $y$, the group velocity $v_{g}=\int dy \ j_{x}^{(n)}(y)$ has a well
defined sign and has been verified from the dispersion curves by means of the
relation $v_{g}=\frac{1}{\hbar}\frac{dE_{n}}{dk_{x}}$. In order to get a better
understanding of the shape of the current density we have plotted together
the electron and the current densities in Fig. \ref{9} and\ref{10}. The density
of
current oscillates between positive and negative ``channels'' as a function of
the electron density and then is a reflection of the quantisation of the
energy.
It is interesting to note that in the study of the FQHE there is also the
appearance of channels which can be seen there as alternating strips of
compressible and incompressible fluid \cite{chk2,ben}. When $B_{0}\neq 0$ the
current density increases with $B_{0}$ and flows in opposite directions on both
sides of the sample when $B_{0}$ is large enough compared to $B_{1}$ or in
other
words when $B_{0}$ is large enough to overcome the effective confining
potential
due to $B_{1}$ so that the electrons are confined by the external confining
potential $V_{c}$. Moreover the current density becomes asymmetric.
\\
Until now only the first term in Eq.(7) has been considered but there is a
second term containing the derivative of the electron density and directly
related to the presence of the spin for the electron. Because, as we have seen
above, the electron density displays a very rich structure, one can expect some
contribution to the current density due to the spin of the electron. This is
shown in Figs. \ref{11} and \ref{12}. Although the part due to the spin is
smaller than the first term in Eq.(7) it is nevertheless noticable. This could
imply some interesting phenomena in relation to spin polarised currents. The
problem is that the ``channels'' are more or less at the same position for spin
up or down which makes it quite difficult to distinguish between both spin
directions.
On the other hand, many parameters can be varied, such as the magnetic field,
the width of the system or the external confining potential which may
allow us to find a suitable system configuration for the production of
polarised
currents.
\\
Finally it has to be stressed that all the discussion above concerned the
current density. Although this quantity is non-zero and has a rich structure,
it
does not imply that the net current $I_{x}=\sum_{k_{x}}\int j_{x}(y) \ dy$ is
non-zero. In fact as our calculation showed $E_{F}$ is independent of $k_{x}$
which means there is no difference of potential across our system and thus no
net current.
\section{Conclusions}
In this work we have studied the effect of a linear magnetic field in a 2DEG.
In
certain simplified cases we were able to carry out some analytical calculations
and to derive the dispersion curves $E_n(k_{x})$ for quite a large range of
$k_{x}\neq 0$. These results were found to be in very good agreement with our
numerical results. In the general case with an external confining potential we
carried out numerical calculations. We derived the whole of the dispersion
curve
and using it, we calculated the electron and current densities. It is
worthwhile
noting that for this calculation we need to consider the states for $k_{x} > 0$
{\it as well as} states for $k_{x} < 0$ and $k_{x} \simeq 0$. This point is
important and could help in understanding  some recent results \cite{kir}
obtained in the framework of CF theory. For the derivation of the electron and
current density we first calculated the Fermi energy $E_{F}$
taking into account that the number of electrons is constant. It turned
out that although $E_{F}$ is independent of $k_{x}$ it is, however, a function
of
the magnetic field. This is due to the external confining potential. The
electron and current densities show a very rich structure which can be seen
as a consequence of the quantisation of the energy, although there is no simple
relation between the $k_{x}$ and $y$ space as is the case for a constant
magnetic field. Moreover the current density exhibits alternating ``channels''
of positive and negative current. It would be interesting too to include
interaction between electrons and to study the effect of the self-consistency
on the way the energy levels cross the Fermi energy. This can give rise to
interesting phenomena, particularly in connection with the shape of the
electron
density \cite{chk2,shi}. Finally, because $\rho(y)$ is not constant we
have a contribution to the current density directly due to the spin of the
electron which could imply some interesting phenomena in relation to spin
polarised currents.
\newpage
\appendix
\section{}
Using the properties of the Hermite polynomial
the matrix elements $H_{kn}$ can be derived in a recursive way.
$\langle\phi_{k}|\left(\frac{p_{\hat{y}}^{2}}{2m'}+a\hat{y}^{4}+
b\hat{y}^{3}+c\hat{y}^{2}+d\hat{y}+e\right)|\phi_{n}\rangle$ gives
\\
\\
$n=k$
\[
\frac{1}{2m'}\left(\frac{1}{2}+k\right)+
3a\left(k\left(1+\frac{(k-1)}{2}\right)+\frac{1}{4}\right)+
c\left(\frac{1}{2}+k\right)+e
\]
$n=k+1$
\[
\left(\frac{(k+1)}{2}\right)^{\frac{1}{2}}
\left(\frac{3b}{2}(1+k)\right)
\]
$n=k+2$
\[
\left(\frac{(k+1)(k+2)}{4}\right)^{\frac{1}{2}}
\left(-\frac{1}{2m'}+a(3+2k)+c\right)
\]
$n=k+3$
\[
\left(\frac{(k+1)(k+2)(k+3)}{8}\right)^{\frac{1}{2}}b
\]
$n=k+4$
\[
\left(\frac{(k+1)(k+2)(k+3)(k+4)}{16}\right)^{\frac{1}{2}}a
\]
\\
and $\langle\phi_{n}|V_{c}|\phi_{m}\rangle$ due to the confining potential
\\
\\
$n=k$
\[
2\:\beta\:{\rm e}^{-\alpha\left(1-\frac{\alpha}{4a^{2}}\right)}\:
L^{0}_{k}\left(-\frac{\alpha^{2}}{2a^{2}}\right)
\]
$n=k+i$
\[
\left(\beta\:{\rm e}^{-\alpha\left(1-\frac{\alpha}{4a^{2}}\right)}\:
\left(\frac{2^{i}}{n(n-1)\dots(n-i+1)}\right)^{\frac{1}{2}}\:
\left(\frac{\alpha}{2a}\right)^{i}\:
L_{k}^{i}\left(-\frac{\alpha^{2}}{2a^{2}}\right)\right)
\left(1+(-1)^{2k+i}\right)
\]
\\
with $L_{n}^{n-k}$ the associate Laguerre polynomial.

\newpage
\begin{figure}
\caption{Dispersion curves obtained analytically compared with numerical
results. $V_{c}=0$, $B_{1}=1\ G/\mbox{{\it \AA}}$ and $B_{0}=0$. The curves
plotted
correspond to the levels $n=0,1,10,11$.}
\label{1}
\end{figure}

\begin{figure}
\caption{Dispersion curves with $V_{c}\neq 0$, $B_{1}=1 \
G/{\it \AA}$ and $B_{0}=0.5 \times 10^{4} \ G$.}
\label{2}
\end{figure}

\begin{figure}
\caption{Fermi energy as a function of the magnetic field ${\bf B}$.}
\label{3}
\end{figure}

\begin{figure}
\caption{Electron densities for different magnetic fields. The solid line
corresponds to the case without an external confining potential.}
\label{4}
\end{figure}

\begin{figure}
\caption{Electron density $\rho(y)$ for
$B_{1}= 1 \ G/{\it \AA}$ and $B_{0}=0$. The insert shows the number of states
as a function of $k_{x}$. The number of oscillations in $\rho(y)$ corresponds
to the number of states.}
\label{5}
\end{figure}

\begin{figure}
\caption{Same as Fig. 5 but with $B_{1}= 0.5 \ G/{\it \AA}$ and
$B_{0}=2\times 10^{4} \ G$.}
\label{6}
\end{figure}

\begin{figure}
\caption{Current density for two different magnetic fields. --- :
$B_{1}= 1 \ G/{\it \AA}$, $B_{0}=0$ and $\cdots$ :
$B_{1}= 0.5 \ G/{\it \AA}$, $B_{0}=2 \times 10^{4} \ G$.
}
\label{7}
\end{figure}

\begin{figure}
\caption{Current density calculated for a fixed $k_{x}$ and a fixed $n$. For
$k_{x}= -0.018 \ {\it \AA}^{-1}$ ($\cdots$), $k_{x}= 0.08 \ {\it \AA}^{-1}$
(-$\cdot$-), and
$n=0$ the movement of the electron is well defined by its classical orbits
(edge and snake orbits). Around $k_{x}=0$, here $k_{x}=0.005 \ {\it \AA}^{-1}$
and
$n=31$ (---),
the situation is more complicated. The density of current can be positive
{\it and} negative as a function of $y$.}
\label{8}
\end{figure}

\begin{figure}
\caption{Total density of current (sum over all states with $E\leq E_{F}$) for
$B_{1}= 0.5 \ G / {\it \AA}$, $B_{0}=2 \times 10^{4} \ G$. The density of
current (---)
oscillates between positive and negative ``channels'' as a function of the
electron density ($\cdots$).
}
\label{9}
\end{figure}

\begin{figure}
\caption{Same as Fig. 9 but with $B_{1}= 3 \ G / {\it \AA}$, $B_{0}=2 \times
10^{4} \ G$.}
\label{10}
\end{figure}

\begin{figure}
\caption{Contribution due to the spin to the current density. The dotted line
is the other contribution in Eq.(7) to the current density.
$B_{1}= 0.5 \ G / {\it \AA}$, $B_{0}=2 \times 10^{4} \ G$.}
\label{11}
\end{figure}

\end{document}